\documentstyle[epsfig]{article}
\begin{document}
\def\unmezzo{{\textstyle 1\over 2}}
\def\dueterzi{{\textstyle 2\over 3}}
\def\O{{\cal O}}
\def\N{{\cal N}}
\def\>t{>_{\scriptscriptstyle{\rm T}}}
\def\kint{\int{\d^3k\over(2\pi)^3}}
\def\pint{\int{\d^3p\over(2\pi)^3}}
\def\gint{\int[\D g]\P[g]}
\def\nbar{\overline n}
\def\barthe{{\bar\theta}}
\def\d{{\rm d}}
\def\e{{\bf e}}
\def\x{{\bf x}}
\def\0x{\x^\smalze}
\def\k{{\bf k}}
\def\X{{\bf X}}
\def\sperpx{{x_\perp}}
\def\sperpk{{k_\perp}}
\def\sbperpk{{{\bf k}_\perp}}
\def\sbperpx{{{\bf x}_\perp}}
\def\perpx{{x_{\rm S}}}
\def\perpk{{k_{\rm S}}}
\def\bperpk{{{\bf k}_{\rm S}}}
\def\bperpx{{{\bf x}_{\rm S}}}
\def\p{{\bf p}}
\def\q{{\bf q}}
\def\zr{{\bf z}}
\def\R{{\bf R}}
\def\A{{\bf A}}
\def\v{{\bf v}}
\def\cm{{\rm cm}}
\def\l{{\bf l}}
\def\sec{{\rm sec}}
\def\Ckol{C_{Kol}}
\def\flux{\bar\epsilon}
\def\zq{{\zeta_q}}
\def\b{b_{kpq}}
\def\bun{b^{\scriptscriptstyle (1)}_{kpq}}
\def\bdu{b^{\scriptscriptstyle (2)}_{kpq}}
\def\z0q{{\zeta^{\scriptscriptstyle{0}}_q}}
\def\smalS{{\scriptscriptstyle {\rm S}}}
\def\smalze{{\scriptscriptstyle (0)}}
\def\smalI{{\scriptscriptstyle {\rm I}}}
\def\smalun{{\scriptscriptstyle (1)}}
\def\smaldu{{\scriptscriptstyle (2)}}
\def\smaltr{{\scriptscriptstyle (3)}}
\def\smalL{{\scriptscriptstyle{\rm L}}}
\def\smalD{{\scriptscriptstyle{\rm D}}}
\def\smal1n{{\scriptscriptstyle (1,n)}}
\def\smaln{{\scriptscriptstyle (n)}}
\def\smalA{{\scriptscriptstyle {\rm A}}}
\def\shell{{\tt S}}
\def\ball{{\tt B}}
\def\nav{\bar N}
\def\micron{\mu{\rm m}}
\font\brm=cmr10 at 24truept
\font\bfm=cmbx10 at 15truept
\centerline{\brm The role of tank-treading motions}
\vskip 3pt
\centerline{\brm in the transverse migration of a}
\centerline{\brm spheroidal vesicle in a shear flow}
\vskip 20pt
\centerline{Piero Olla}
\vskip 5pt
\centerline{CRS4 via N. Sauro 10}
\centerline{09123 Cagliari Italy}
\centerline{and}
\centerline{ISIAtA-CNR}
\centerline{Universit\'a di Lecce}
\centerline{73100 Lecce Italy}
\vskip 20pt
\centerline{\bf Abstract}
The behavior of a spheroidal vesicle, in a plane shear flow bounded from one 
side by a wall, is analysed when the distance from the wall is much larger than 
the spheroid radius. It is found that tank treading motions produce a transverse
drift away from the wall, proportional to the spheroid eccentricity and the 
inverse square of the distance from the wall. This drift is independent of 
inertia, and is completely determined by the characteristics of the vesicle 
membrane. The relative strength of the contribution to drift from tank-treading 
motions and from the presence of inertial corrections, is discussed. 
\vskip 2cm
\centerline{\it J. Phys. A (Math. Gen.)}
\centerline{\it In press}
\vfill\eject

\centerline{\bf I. Introduction}
\vskip 5pt
An important topic in the study of suspensions is to understand the ability of 
shear to induce particle transport perpendicular to the velocity lines
\cite{segre62,segre62a}. 
One of the mechanisms responsible for this kind of transport is the presence of
lift forces due to inertia, pushing the particles away from the walls \cite{saff65} 
(for recent references, see \cite{mclaugh93}). It is well
known that the symmetry properties of the Stokes equation, which governs the 
dynamics of a purely viscous fluid, do not allow
for the presence of lift perpendicular to the flow lines, at least for the case
of spherical particles suspended in a plane shear flow \cite{brether62}. However, 
mechanisms different from inertia may allow for the symmetry breaking, necessary for the 
production of a transverse drift.

A mechanism that is important, if one deals with a suspension of deformable objects,
is the ability of shear to induce, under appropriate circumstances, a fixed
orientation of the particles in suspension. If the deformable objects are vesicles 
filled with some other fluid, like in the case of blood, red cells, a fixed 
orientation can be attained when the vesicles are in a state of tank-treading 
motion \cite{keller82}. In this state, the membrane and the fluid inside circulate in 
a steady fashion around the cell interior, while the cell shape and orientation
remain constant.

Taking a comoving cartesian system $\{ x_1,x_2,x_3\}$, with origin at the cell 
centre, and oriented in such a way that the unperturbed velocity field has 
components: $\bar\v=(0,0,\alpha x_2)$, there will be a velocity perturbation $\v$
such as, if the cell is exactly spherical and inertia is neglible:
$v_{1,2}(x_3)=-v_{1,2}(-x_3)$. Under 
these conditions, a plane wall perpendicular to the $x_2$ axis, will produce a
correction $\v^\smalI$ to the perturbation field, such as, by symmetry $v_2^\smalI=0$ at 
the cell
centre. An ellipsoidal cell with fixed orientation, instead, will lead to symmetry 
breaking. Hence, there will be a drift $v^\smalL=v_2^\smalI(0)$, which will depend on the 
cell eccentricity, on the boundary conditions at the cell surface and, of course,
on the distance from the wall $l$. Notice however that a non spherical shape is not by 
itself  a sufficient condition for lift. A rigid non-spherical object for instance, 
will not  keep an orientation fixed in time, carrying on instead a kind of flipping 
motion \cite{jeffery22}, whose contribution to lift is going to be much reduced. 

Inertia does not play any role in the mechanism for lift outlined above,
and it will appear that its effect can be disregarded, when the particle is 
sufficiently close to the wall. Actually, depending on the circumstances, the
contribution to drift from tank treading motions may even result to be dominant. 
The drift of a spherical particle in a bounded shear flow, is purely due to inertia; 
in the case of a particle in a Couette gap with thickness $L$:
$v^\smalL(l)=f(l/L) Re_p\alpha R$, where $R$ is the particle radius, 
$Re_p={\alpha R^2\over\nu}$, with $\nu$ the kinematic viscosity of the external fluid, 
is the particle Reynolds number, and $f(l/L)$ is at most of the order of a few tenths
\cite{vasseur76}. In the case of a tank-treading motion generated drift, one has 
instead, in place of $Re_p$, some parameter describing the non-sphericity of the 
particle, which, like in the case of red cells \cite{oiknine83}, is not necessarily small.

In this paper, the simplest possible case of a neutrally buoyant, almost spherical
vesicle in a purely Newtonian solvent, is taken into consideration. This allows a 
perturbative analysis around the standard case of a spherical particle.
The vesicle is taken at a distance $l$ from the wall such as: 
$R\ll l\ll\sqrt{\nu/\alpha}$; in this range, it is possible at the same time to 
neglect inertia, and to treat the effect of the wall, as a correction to the 
velocity perturbation due to the vesicle. The technique is similar to the one adopted
by Ho and Leal \cite{ho74} in the case of a spherical particle in a bounded shear flow. 

The full problem of the determination of the spheroid shape under the combined effect
of the stresses in the solvent, the membrane and the fluid inside is not treated (for
reference about this problem, see  e.g. \cite{pozri90,barthes80}).
Rather, an axisymmetric, ellipsoidal vesicle shape is assumed. Thus, no information 
on the nature of the membrane and of the fluid inside is utilised. However,
two kinds of motions at the vesicle boundary are analysed: one which is area 
preserving, mimicking the behavior of an inextensible membrane, like that of blood 
cells, and one in which points at the surface move with uniform angular velocity, 
simulating the case of an immaterial interface.


\vskip 10pt
\centerline{\bf II Stokes equation in vector spherical harmonics}
\vskip 5pt
In a perturbative analysis around a spherically symmetric situation, it is worth while
expanding the velocity field in terms of vector spherical harmonics:
$$
\v(\x)=\v^{\rm s}(\x)+\v^{\rm e}(\x)+\v^{\rm m}(\x)
$$
$$
=\sum_{lm}[v^{\rm s}_{lm}(x)\x Y_{lm}(\e_x)+
	 v^{\rm e}_{lm}(x)\nabla Y_{lm}(\e_x)+
	 v^{\rm m}_{lm}(x)[\x\times\nabla]Y_{lm}(\e_x)],
\eqno(1)
$$
where $Y_{lm}(\e_x)$ are standard, normalized spherical harmonics and $\e_x=\x/x$; the 
superscripts $\{ {\rm sem}\}$ stand for scalar, electric and magnetic and come from
the origin of this basis as a tool in the study of electromagnetic waves \cite{landau}. 

This basis is clearly orthogonal and the components $v^{\rm sem}_{lm}$ can be obtained
in the standard way. Alternatively, one can write:
$$
v^{\rm s}_{lm}={<lm|\x\cdot\v>\over x^2},
\eqno(2)
$$
$$
v^{\rm e}_{lm}=-{x^2\over l(l+1)}<lm|\nabla_\perp\cdot(\v-\v^{\rm s})>
\eqno(3)
$$
and
$$
v^{\rm m}_{lm}={<lm|\nabla\cdot[\x\times\v^{\rm m}]>\over l(l+1)},
\eqno(4)
$$
where the bra-ket notation  
$<lm|f>\equiv f_{lm}(x)=\int\d\Omega_xY^*_{lm}(\e_x)f(\x)$, with  
$\d\Omega_x$  the solid angle differential, is used, and $\nabla_\perp$ is the angular part
of the gradient.

At stationarity, an incompressible fluid in creeping flow conditions, obeys the
time independent linearized version of the vorticity equation:
$$
\nabla^2[\nabla\times\v]=0
\eqno(5)
$$
together with the continuity equation:
$$
\quad\nabla\cdot\v=0.
\eqno(6)
$$
In terms of $v^{\rm sem}_{lm}$ components, the vorticity equation reads:
$$
\Big({\d^2\over\d x^2}+{2\over x}{\d\over\d x}-{l(l+1)\over x^2}\Big)f^{(1,2)}_{lm}=0
\eqno(7)
$$
where:
$$
f^{(1)}_{lm}=-v^{\rm s}_{lm}+{1\over x}v^{\rm e}_{lm}\quad{\rm and}\quad
f^{(2)}_{lm}=-v^{\rm m}_{lm},
\eqno(8)
$$
while the continuity equation takes the form:
$$
x{\d v^{\rm s}_{lm}\over\d x}+3v^{\rm s}_{lm}-{l(l+1)\over x^2}v^{\rm e}_{lm}=0
\eqno(9)
$$
From Eqns. (7-9), one obtains the ''outside'' and ''inside'' solutions:
$$
\cases{v^{\rm s}_{lm}=a_{lm}x^{-1-l}+b_{lm}x^{-3-l},\cr
v^{\rm e}_{lm}={2-l\over l(l+1)}a_{lm}x^{1-l}-{b_{lm}\over l+1}x^{-1-l},\cr
v^{\rm m}_{lm}=c_{lm}x^{-1-l};\cr}
\eqno(10)
$$
and:
$$
\cases{v^{\rm s}_{lm}=a_{lm}'x^l+b_{lm}'x^{l-2},\cr
v^{\rm e}_{lm}={l+3\over l(l+1)}a_{lm}'x^{l+2}-{b_{lm}'\over l}x^l,\cr
v^{\rm m}_{lm}=c_{lm}'x^l.\cr}
\eqno(11)
$$
The expression for the velocity perturbation by a spherical particle in a strain flow,
is thus obtained. Remembering the correspondence between fully symmetric, traceless
$l$-tensors and spherical harmonics with given $l$, the contraction:
$x^{-2}\alpha_{ij}x_ix_j$, where $\alpha_{ij}=\alpha_{ji}$ and $\alpha_{ii}=0$ (the
convention of summation over repeated indices is adopted) can be expressed as a linear
combination of spherical harmonics with $l=2$. One finds then from Eqn. (10): 
$$
v_i=\Big({7\over 6}ax^{-5}+{5\over 3}bx^{-7}\Big)x_i\alpha_{jk}x_jx_k-{2\over 3}b
x^{-5}\alpha_{ij}x_j,
\eqno(12) 
$$
which is, for $a=-{10b\over 7}$ the velocity perturbation due to a sphere of unitary
radius put in a strain flow: $\bar v_i={2b\over 3}\alpha_{ij}x_j$ \cite{batch}.


\vfill\eject
\centerline{\bf III. Velocity perturbation by an ellipsoidal vesicle}
\vskip 5pt
\noindent{\it Boundary conditions at the wall}
\vskip 5pt
An ellipsoidal cell in a shear flow will feel the effect of the strain component of the 
flow, which will tend to align strain and ellipsoid axes, while
the vortical component will tend to make the cell rotate. If the
viscosity of the fluid inside or the rigidity of the membrane are too large, the vesicle
will tend to rotate as a rigid body, otherwise, it will be in a state of tank-treading 
motion. 

\begin{figure}[hbtp]\centering
\centerline{
\epsfig{figure=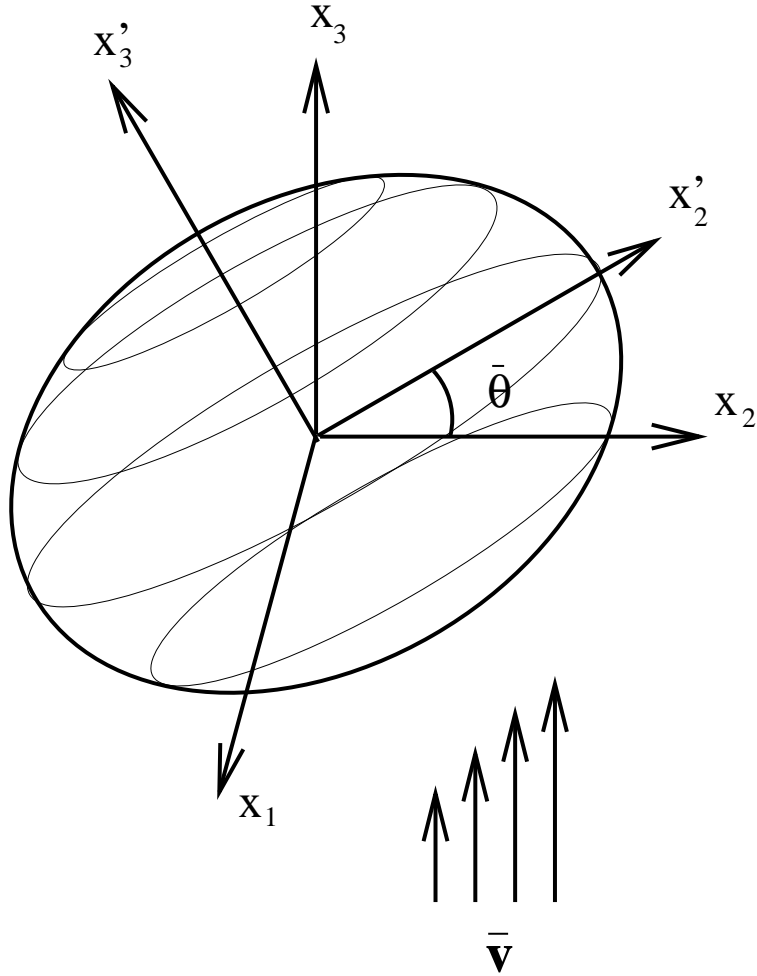,width=10.cm}
 }
\caption{Cell orientation in a plane  shear flow $\bar\v=[0,0,\alpha x_2]$}
\end{figure}

\noindent In this configuration the symmetry axis of the ellipsoid will lie in the plane
$x_2x_3$, at an angle $\bar\theta$ with respect to the $x_2$, with $\barthe\to\pm\pi/4$ 
for vanishing resistance of the vesicle to deformation, and $\barthe\to\pm\pi/2$ at the
threshold for the transition to flipping motion \cite{keller82}. In the two limits, the
long side of the ellipsoid tends to be parallel, respectively, to the expanding direction
of the strain and to the unperturbed velocity field.

Taking a reference system 
$\{ x_1'x_2'x_3'\}$ at the cell centre, as shown in Fig. 1, with $x_3'$ along 
the symmetry axis of the ellipsoid and 
$x_1'\equiv x_1$, the equation for the cell surface can be written in the form:
$$
(1-\epsilon)(x_1'^2+x_2'^2)+(1+\epsilon)x_3'^2=R^2
\eqno(13)
$$
For small values of the eccentricity $\epsilon$, the distance, from the origin of a point 
on the surface, with given elevation $x'_3$ (and corresponding angular coordinates in the
unprimed frame $\theta$ and $\phi$), can be written in the form:
$$
x(\theta,\phi)\simeq
R\Big(1+{\epsilon\over 2}\Big(1-2\Big({x_3'\over R}\Big)^2\Big)\Big)
$$
$$
=R\Big(1+{\epsilon\over 2}\Big(1-2{(\cos\barthe x_3-\sin\barthe x_2)^2\over R^2}\Big)\Big), 
\eqno(14)
$$
In a state of tank-treading motion, the membrane is assumed to move following two 
possible laws. In the first case, for each value of $x_1$, points on the surface are 
taken to move at constant speed: 
$$
v_B={1\over 2}B\alpha R(x_1),\quad{\rm with:}\quad R(x_1)^2=R^2
-(1-\epsilon)x_1^2,\eqno(15)
$$
remaining on the elliptic trajectory, which corresponds to the given section at constant $x_1$ 
of the cell surface. Such motion is locally area preserving, and approximates the behavior of
an inextensible membrane, with negligible resistance to bending and shear. For small $\epsilon$, 
one expects $B\simeq 1+\beta\epsilon$; thus, for $\epsilon=0$, the expression for $v_B$ 
reduces to that for a sphere immersed in the given shear flow. In terms of components:
$$
\v_B=v_B\Big(0,{x'_3\over\sqrt{{x'}_3^2+(\gamma x'_2)^2}},
{x'_2\over\sqrt{{x'}_2^2+(x'_3/\gamma)^2}}\Big)
\eqno(16)
$$
where $\gamma=\sqrt{1-\epsilon\over 1+\epsilon}$. For small $\epsilon$:
$$
\v_B\simeq{\alpha\over 2}[0,-x_3'(1+\epsilon (\beta+{x_1^2\over 2x_\perp^2}+
{2{x'_2}^2\over x_\perp^2})),x'_2(1+\epsilon (\beta+{x_1^2\over 2x_\perp^2}-
{2{x'_3}^2\over x_\perp^2}))]
\eqno(17)
$$
where $x_\perp^2={x'_2}^2+{x'_3}^2$. Using Eqns. (2) and (17), one finds:
$$
\v_B^{\rm s}\simeq-\epsilon\alpha\x{x'_2x_3'\over x^2}\eqno(18)
$$
and:
$$
\v_B^{\rm e}\simeq\epsilon\alpha
\Big({\x\over x^2}-{\x_\perp\over x_\perp^2}\Big)x'_2x_3'
\eqno(19)
$$
It will appear that only the $l=2$ part of $\v_B$ must be taken in consideration to 
calculate the lift. Using Eqn. (3), a simple calculation leads to the 
result:
$$
\v_B=\v_B^{\rm m}-{5\over 6}\epsilon\alpha\x{x_2'x_3'\over x^2}
-{\epsilon\alpha\over 12}[0,x_3',x_2']+H.H.,
\eqno(20)
$$
where $H.H.$ indicates higher harmonics. In terms of unprimed variables and components:
$$
\v_B=\v_B^{\rm m}-{5\over 12}\epsilon\alpha\x{(x_3^2-x_2^2)\sin 2\barthe
+2x_2x_3\cos 2\barthe\over x^2}
$$
$$
-{\epsilon\alpha\over 12}[0,-x_2\sin 2\barthe+x_3\cos 2\barthe,
x_3\sin 2\barthe+x_2\cos 2\barthe]+H.H.
\eqno(21)
$$
The second membrane behavior that is considered, is characterized by a constant angular
velocity motion:
$$
\v_B={B\alpha\over 2}[0,-x_3/\gamma,\gamma x_2]\simeq{\alpha\over 2}(1+\beta\epsilon)[0,-x'_3,x_2']
-{\alpha\epsilon\over 2}[0,x'_3,x'_2],\eqno(22)
$$
In terms of unprimed variables:
$$
\v_B\simeq{\alpha\over 2}(1+\beta\epsilon)[0,-x_3,x_2]
$$
$$
-{\alpha\epsilon\over 2}[0,-x_2\sin 2\barthe+x_3\cos 2\barthe,
x_3\sin 2\barthe+x_2\cos 2\barthe].
\eqno(23)
$$ 
The behavior described in Eqns. (22) and (23) should mimic the motion of a droplet; 
however, this is also the approximation for an area preserving membrane motion, 
used by Keller and Skalak in \cite{keller82}. Thus, comparing the velocity perturbation
and the lift produced by the two membrane motions, prescribed in  Eqns. (17) and
(23), gives, among the other things, an idea on the acceptability of such an approximation,
also in view of the simplifications that become possible in the large $\epsilon$ regime.

\vskip 8pt
\noindent{\it Calculation of the perturbation in an infinite fluid}
\vskip 5pt

If the membrane velocity $\v_B(\theta,\phi)$ and the cell orientation $\barthe$, 
produced by the external flow $\bar\v(\theta,\phi,x)$, are supposed known, the velocity 
perturbation $\v$ can be obtained from the boundary condition at the vesicle surface 
$x=x(\theta,\phi)$:
$$
\v_B(\theta,\phi)=\bar\v(\theta,\phi,x(\theta,\phi))+\v(\theta,\phi,x(\theta,\phi)).
\eqno(24)
$$ 
For small values of $\epsilon$, it is possible to solve Eqn. (24) perturbatively; to 
$\O(\epsilon)$:
$$
\v_B^\smalze(\theta,\phi)=\bar\v(\theta,\phi,R)+\v^\smalze(\theta,\phi,R)
\eqno(25{\rm a})
$$
$$
\v_B^\smalun(\theta,\phi)=\bar\v(\theta,\phi,x(\theta,\phi))+
\v^\smalze(\theta,\phi,x(\theta,\phi))+\v^\smalun(\theta,\phi,R),
\eqno(25{\rm b})
$$
where, for $x(\theta,\phi)$, the $\O(\epsilon)$ correct expression, provided by Eqn. (14)
can be used. In Eqn. (25a) $\v_B^\smalze$, is the velocity at the surface of the sphere, 
with $x^\smalze (\theta,\phi)=R$, immersed in the same flow:
$$
\v_B^\smalze(\theta,\phi)={\alpha\over 2}[0,-\0x_3,\0x_2];
\eqno(26)
$$ 
In this way, $\v^\smalze$ is the velocity perturbation by a sphere in a shear flow, given
by Eqn. (12):
$$
\v^\smalze={5\over 2}\Big({R^4\over x^4}-{R^2\over x^2}\Big){\alpha x_2x_3\x\over x^2}
-{\alpha R^4\over x^4}[0,x_3,x_2]
\eqno(27)
$$
Expanding Eqn. (25b) in vector spherical harmonics and using Eqn. (10), allows then to 
calculate $\v^\smalun$ for $x>x(\theta,\phi)$. 

For the first membrane motion, described by Eqn. (21), one has:
$$
\v_B^\smalun={\alpha\over 2}\Big((\beta+{x_1^2\over 2x_\perp^2})\epsilon+
{\delta x\over R}\Big)[0,-x_3,x_2]+\v_C\eqno(28)
$$
where:
$$
\v_C=-{5\over 12}\epsilon\alpha\x{(x_3^2-x_2^2)\sin 2\barthe
+2x_2x_3\cos 2\barthe\over R^2}
$$
$$
-{\epsilon\alpha\over 12}[0,-x_2\sin 2\barthe+x_3\cos 2\barthe,
x_3\sin 2\barthe+x_2\cos 2\barthe]+H.H.
\eqno(29)
$$
In Eqns. (28) and (29):
$$
\delta x={R\epsilon\over 2}\Big(1-2{(\cos\barthe x_3-\sin\barthe x_2)^2\over R^2}\Big), 
\eqno(29)
$$
and the identification ${\bf \0x}=\x$ is allowed at the order in $\epsilon$ considered.
It appears (details are given in the appendix) that the condition on the absence of
external torque on the vesicle, corresponds to the $l=1$ components of $\v^{\rm m}$, being
identically zero. This is achieved choosing appropriately the coefficient $\beta$, and 
consequently, the value of the tank treading velocity.
The present analysis is directed to calculate the lift at large distance from the wall. 
Hence, from Eqn. (10), the leading contribution is from $\v^{\rm s}_{l=2}$ terms. 
Substituting Eqns. (14), (27) and (28) into Eqn. (25b), keeping only $\O(\epsilon)$
terms, and neglecting $\v^{\rm m}$ and $l>2$ components, one obtains for $\v^\smalun$:
$$
\v^\smalun_{x=R}=\v_C+{5\alpha\delta x\over R}\Big( {x_2x_3\x\over R^2}
-{1\over 2}[0,x_3,x_2]\Big).
\eqno(30) 
$$
Using Eqn. (29) together with Eqns. (2-3), one finds:
$$
\v^{\smalun\rm s}_{x=R}=-{\epsilon\alpha\x\over 2R^2}
[(x_3^2-x_2^2)\sin 2\barthe+2x_2x_3\cos 2\barthe]
\eqno(31)
$$ 
and:
$$
\v^{\smalun\rm e}_{x=R}\simeq-\epsilon\alpha R^2\nabla
{(x_3^2-x_2^2)\sin 2\barthe+(2\cos 2\barthe+30)x_2x_3\over x^2}
$$
$$
-{5\epsilon\alpha\over R^2}(\cos\barthe x_3-\sin\barthe x_2)^2
\Big({x_2x_3\x\over R^2}-{1\over 2}[0,x_3,x_2]\Big).
\eqno(32)
$$
All terms in Eqns. (31) and (32) are clearly $l=2$ with the exception of the last line
of Eqn. (32), which is a combination of $\v^{\rm e}$ and $\v^{\rm m}$ terms. After
some tedious calculations, expedited by the use of Eqn. (3), the following expression 
for the $l=2$ part of $\v^{\smalun\rm e}_{x=R}$, is obtained:
$$
\v^{\smalun\rm e}_{x=R}=
\epsilon\alpha R^2\nabla\Big({\sin 2\barthe\over x^2}\Big({5x_1^2\over 42}-{x_2^2\over 56}
-{17x_3^2\over 168}\Big)
-\Big({25\over 84}+{\cos 2\barthe\over 12}\Big){x_2x_3\over x^2}\Big)
\eqno(33)
$$
The leading term at large distance is, in the notation of Eqn. (10), the contribution
$a_{2m}x^{-3}$ to $v^{\rm s}$. From Eqns. (31) and (33), using Eqn. (10), one obtains
therefore the large distance result:
$$
\v^\smalun\simeq{\epsilon\alpha R^3\x\over x^3}
\Big({\sin 2\barthe\over x^2}\Big({5x_1^2\over 14}+{25x_2^2\over 56}
-{45x_3^2\over 56}\Big)
-\Big({25\over 28}+{5\cos 2\barthe\over 4}\Big){x_2x_3\over x^2}\Big).
\eqno(34)
$$
The same identical calculations can be carried on for the second kind of membrane motion 
described by Eqn. (23), with the final result:
$$
\v^\smalun\simeq{\epsilon\alpha R^3\x\over x^3}
\Big({\sin 2\barthe\over x^2}\Big({5x_1^2\over 14}+{15x_2^2\over 14}
-{10x_3^2\over 7}\Big)
-\Big({25\over 28}+{5\cos 2\barthe\over 2}\Big){x_2x_3\over x^2}\Big).
\eqno(35)
$$
One question that comes natural at this point is whether one could extend the analysis to
a non perturbative regime, expanding directly Eqn. (24) into vector spherical harmonics and 
then, after imposing a cut-off on $l$ and $m$, solving numerically the resulting linear 
system in the unknowns $a_{lm}$, $b_{lm}$ and $c_{lm}$. This could be useful to analyse the 
behavior of vesicles of a general, strongly non spherical shape, but it is not the 
most efficient way of proceeding, also because the matrix associated with the system is 
strongly ill conditioned. Of course, in the case of an ellipsoid immersed in a shear flow, 
undergoing a linear tank-treading motion like the one of Eqn. (22), the theory of Keller 
and Skalak \cite{keller82} can be utilised to  obtain analytic expressions for the velocity 
field. In the more general case, it turns out that it is possible to modify the initial vector
spherical harmonics basis, in such a way that the velocity field is obtained directly,
without having to invert any ill conditioned matrices \cite{olla}.


\vskip 10pt
\centerline{\bf IV. Calculation of lift}
\vskip 5pt
The calculations in the previous section referred to an unbounded flow situation. A
wall at a large distance from the vesicle will cause a correction $\v^\smalI$ to 
the velocity perturbation $\v$ obtained in the previous section. This can be calculated
using the boundary condition provided by making the correction at the wall, equal to 
minus the velocity given by Eqns. (34) or (35). The higher order corrections
are obtained then, using the value, alternately on the vesicle surface and on the
wall, of the previous calculated correction as boundary condition, in a series of images and 
counter-images, analogous to those of electrostatics. If the distance from 
the wall is large, it is possible however to stop at the first image. 

The technique at this point is standard \cite{ho74, quarta}; one introduces scalar and vector 
potential $\phi$ and $\A$, such as:
$$
\v^\smalI=\nabla\phi+\nabla\times\A
\eqno(36)
$$
where:
$$
\nabla^2\phi=0\quad{\rm and}\quad\nabla\cdot\A=0.
\eqno(37)
$$
The first of Eqn. (37) is a consequence of incompressibility, while the second is a gauge
condition. From here, the vorticity equation takes the form of a biquadratic:
$$
\nabla^2\nabla^2\A=0
\eqno(38)
$$
Taking the wall  at $x_2=l$ positive, parallel to the $x_{1,3}$ axes, 
it is useful to Fourier transform all quantities with respect to $x_1$ and $x_3$:
$f(\x)=\int{\d k_1\d k_3\over (2\pi)^2}\,e^{i(k_1x_1+k_3x_3)}$ $\times\tilde f(k_1,k_3,x_2)$. 
The gauge condition on $\A$ takes then the form:
$$
\tilde A_3=-{k_1\over k_3}\tilde A_1+{i\over k_3}\tilde A'_2,
\eqno(39)
$$
where the prime indicates derivative with respect to $x_2$. The gauge is definitely fixed
by requiring that $\A$ does not contain any potential contribution ${\bf a}$ such as 
$\nabla^2{\bf a}=0$. With this condition, solution of Eqn. (38) gives:
$$
\tilde \A=\hat\A(k_1,k_3)(x_2-l)\exp(k(x_2-l)),
\eqno(40)
$$
where $k=\sqrt{k_1^2+k_3^2}$. The first of Eqn. (37), instead, gives for $\phi$:
$$
\tilde \phi=\hat\phi(k_1,k_3)\exp(k(x_2-l)).
\eqno(41)
$$
Using Eqns. (36) and (39), the expression for the velocity correction becomes, in terms
of Fourier components:
$$
\cases{\tilde v^\smalI_1=-{k_1\over k_3}\tilde A_1'+{i\over k_3}\tilde A_2''-ik_3
		  \tilde A_2+ik_1\tilde\phi,\cr
\tilde v^\smalI_2={ik^2\over k_3}\tilde A_1+{k_1\over k_3}\tilde A_2'+\tilde\phi',\cr 
\tilde v^\smalI_3=ik_1\tilde A_2-\tilde A_1'+ik_3\tilde\phi,\cr}
\eqno(42)
$$
and imposing the boundary condition $\tilde\v^\smalI(k_1,k_3,l)=-\tilde\v(k_1,k_3,l)$, one finds:
$$
\cases{\hat v_1={k_1\over k_3}\hat A_1-{2ik\over k_3}\hat A_2-ik_1\hat\phi,\cr
\hat v_2=-{k_1\over k_3}\hat A_2-k\hat \phi,\cr 
\hat v_3=\hat A_1-ik_3\hat\phi,\cr}
\eqno(43)
$$
where $\hat\v(k_1,k_3)=\tilde\v(k_1,k_3,l)$. Solution of this system gives:
$$
\cases{\hat\phi={i[-k_1k_3\hat v_1+2ik_3k\hat v_2-k_1^2\hat v_3]\over 2k_3k^2},\cr
\hat A_1={k_1k_3\hat v_1-2ik_3k\hat v_2-(k^2+k_3^2)\hat v_3\over 2k^2},\cr 
\hat A_2={i(k_3\hat v_1-k_1\hat v_3)\over 2k}.\cr}
\eqno(44)
$$
At this point, one finds $\tilde v^\smalI_2$, from substitution of Eqn. (44) into (42); for
$x_2=0$:
$$
\tilde v^\smalI_2(k_1,k_3,0)=-[ik_1l\hat v_1+(1+kl)\hat v_2+ik_3l\hat v_3]\exp(-kl)
\eqno(45)
$$
The lift is found from inverse Fourier transform at $x_1=x_3=0$: $v^\smalL(l)=
\int{\d^2k\over(2\pi)^2}v_2^\smalI(k_1,k_3,0)$. Of all the contributions to $\v$ entering 
Eqn. (45), only those proportional to $\sin 2\barthe$ in $\v^\smalun$ give a nonzero 
result. Writing in an  explicit way:
$$
v^\smalL(l)=-{1\over(2\pi)^2}\int_0^\infty k\d k\int_0^\infty\hat x\d\hat x
\int_0^{2\pi}\d\varphi\int_0^{2\pi}\d\psi
\exp(-ik\hat x\cos\psi-kl)
$$
$$
\times [klv_1\cos\varphi+(1+kl)v_2+iklv_3\sin\varphi]
\eqno(46)
$$
where $\hat x=\sqrt{x_1^2+x_3^2}$, $x_1=\hat x\cos\theta$, $k_1=k\cos\varphi$ and 
$\psi=\theta-\varphi$. In Eqn. (46), one can write, from Eqns. (34) and (35):
$$
\v={\epsilon\alpha R^3\sin 2\barthe\over\hat x^5}(2Cl^2+2(D-C)\hat x^2\cos 2\barthe
-D\hat x)[\hat x\cos\theta,l,\hat x\sin\theta],
\eqno(47)
$$
where all terms giving zero contribution in the integrals of Eqn. (46) are disregarded. 
The integrals in Eqn. (46) can be carried on analytically exploiting the following 
properties of the Bessel functions $J_\nu(x)$ \cite{grad}:
$$
\int_0^{2\pi}\d\psi\,\exp(-i\alpha\cos\psi)=2\pi J_0(\alpha)
\eqno(48)
$$
$$
\int_0^{2\pi}\d\psi\,\cos\psi\exp(-i\alpha\cos\psi)=2\pi i J_1(\alpha);
\eqno(49)
$$
and:
$$
\int_0^\infty x^\mu\d x J_\nu(\beta x)\exp(-\alpha x)=(-1)^\mu{\d^\mu\over\d\alpha^\mu}
{(\sqrt{\alpha^2+\beta^2}-\alpha)^\nu\over\beta^\nu\sqrt{\alpha^2+\beta^2}}.
\eqno(50)
$$
The final result is:
$$
v^\smalL=-{\epsilon\alpha C\, R^3\sin 2\barthe\over l^2}.
\eqno(51)
$$
Thus, the magnitude and the direction of the lift depend on the coefficient of the $x_2^2$ 
terms in the expression for $\v^\smalun$ given by Eqns. (34) and (35). It is known 
\cite{keller82}, and a
simple argument is given in the appendix, that for sufficiently small values
of $\epsilon$, $\barthe=\pm{\pi\over 4}$, with the plus sign when the ellipsoid is oblate, 
($\epsilon>0$) and minus when this is prolate ($\epsilon<0$). This leads then, both in the
inextensible membrane case of Eqn. (34), and in the ''droplet'' case of Eqn. 
(35), a drift away from the wall, which is directly proportional to the inverse square of 
the distance from the wall, and to the eccentricity $\epsilon$:
$$
v^\smalL=-{C|\epsilon|\alpha R^3\over l^2}.
\eqno(52)
$$
The values of the constant $C$ are obtained form Eqns. (34-35); one finds in the two cases,
respectively: $C={25\over 112}$ and $C={15\over 28}$.


\vskip 10pt
\centerline{\bf V. Conclusions}
\vskip 5pt
The main result that has been obtained in this paper, is that tank-treading motions 
are able to produce a transverse drift of vesicles in a sheared suspension. 
This effect is mainly localized near the walls, where it can dominate that
of the inertial corrections, on which current theories on the lift of particles in 
suspension are based. 
It is possible to estimate the thickness of the region where this happens. Using 
the expression for the inertial drift of a spherical particle, in a bounded shear
flow with $0<l<L$: 
$f(l/L)\sim f^0\, (l/L-0.5)$, where $f^0$ is a constant of the order of a few tenth 
\cite{vasseur76}, one finds from Eqn. (51), that the drift from tank treading motions 
remains dominant as long as:
$$
{l\over R}<\Big({C\epsilon L\over f^0Re_p R}\Big)^{1\over 3}.
\eqno(53)
$$
Thus, thanks to the smallness in most situations, of the particle Reynolds number $Re_p$, 
the width of the region where the drift from tank-treading motions is dominant can be
so large that both approximations of large $l/R$ and small $\epsilon$, used in this 
paper, can be satisfied at the same time.

However, the most interesting situation is that of strongly non spherical vesicles, for
which $\epsilon=\O(1)$.
It is known that, for large values of $\epsilon$, the vesicle tends to have its major
axis aligned with the unperturbed flow (which corresponds to $\sin 2\barthe=0$), 
and then to make the transition to flipping motion \cite{keller82}. Thus, there must 
be some critical $\bar\epsilon$, for which, fixed the other parameters, the drift 
velocity achieves its maximum value. 

In the present theory, due to the smallness of $\epsilon$, one has:
$\barthe\simeq\pm\pi/4$, and the effect of destruction of tank treading motions, 
produced by $\epsilon$ becoming too large, is not accounted for. At the same time, 
the perturbative calculation leading to the large distance behavior of the velocity 
disturbance $\v$, becomes unreliable. However, extrapolating Eqn. (53) to large values 
of $\epsilon$, the effect of tank-treading motions (when present) is expected to become 
dominant in all situations of suspensions flowing in narrow gaps. Thus a nonperturbative
extension of the theory at large $\epsilon$ would be advisable. 

Red cells are a physical system in which such a non perturbative theory could be applied. However,
experimental observations \cite{fisher80,tran84} as well as theoretical analysis \cite{keller82} 
indicate that, due to their membrane viscosity and that of the hemoglobin inside, these cells
undergo tank treading motions only when immersed in very viscous solvents, or in the presence of 
very strong shear stresses. It is clear on the other hand, that inertia cannot 
account by itself for phenomena like the concentration 
of red cells near the axis of small blood vessels: the so called  Fahraeus-Lindqvist effect 
\cite{oiknine76}. Using typical parameters for a red cell and a small vessel \cite{oiknine83}: 
$R\sim 4\mu$, $L\sim 100\mu$, $\alpha\sim 200\sec^{-1}$ (i.e. $\bar v\sim 1\cm/\sec$), one 
finds that a transverse drift of the order of $L$, would occur only after the cell has travelled 
already for several centimeters, much more than the length of a typical small vessel. Thus, 
a fixed orientation and strong departure from sphericity remain the necessary ingredients for 
a sufficiently large lift, and some different mechanism for maintaining a fixed orientation 
in the cell must be found.

The most important limitation of an analysis like the one in this paper, is the 
fact that the membrane motion and the cell shape are imposed from the outside.
This, among the other things, makes impossible to analyse the behavior of the
cell, when the distance from the wall becomes comparable to the radius $R$.
At the same time, there remain open several questions on which kind of membrane 
motions and cell shapes are possible, for instance whether there could be 
membranes, whose response to an external shear field leads to a drift towards the 
nearest wall, instead of away from it. The two behaviors considered in this work lead 
to lifts of different magnitude, but both directed away from the wall. 

The fact that different membrane behaviors lead to the cell moving in one direction or
another suggests the possibility of controlling the cell motion through the
membrane stiffness and the internal fluid viscosity. This could be of great importance
in microsurgery, where new methods to deliver drugs using microcapsules that break and drop
their content when they are on the selected target, are taken into consideration. Already now, 
it seems realistic to have cells whose stiffness varies with temperature and whose tendency 
to stay away from the walls could be controlled from the outside. Carefully 
designing the cell structure (and its response to external stresses or other stimuli), 
could allow in principle, a fuller control on the cell trajectory, without having to 
rely on internal motors and sources of energy.

\vskip 10pt
\noindent{\bf Aknowledgements}: I would like to thank Pasquale Franzese, Fabio Maggio,
Stephane Zaleski and Gianluigi Zanetti for interesting and helpful conversation. This
research was funded in part by a grant from Regione Sardegna.


\vskip 15pt
\centerline{\bf Appendix: Torque and energy balance}
\vskip 5pt
In a steady state of tank trading motion, a vesicle in a shear flow will feel no external 
torque, while the external work will be equal to the energy dissipated in the membrane
and the cell interior. In this paper, the membrane is supposed to oppose no resistance to
bending and to shear, and consequently, to be dissipation free.
The simultaneous satisfaction of these two conditions fixes the 
magnitude of the tank-treading velocity $v_B$ and the orientation angle $\barthe$.

The external torque is calculated from the value of the stress tensor $T_{ij}$ at a
spherical surface enclosing the vesicle: 
$$
T_{ij}=\mu\nu (\partial_iv_j+\partial_jv_i),
\eqno({\rm A}1)
$$
where $\mu$ is the fluid density.
The force and the torque on an infinitesimal element $\d S_i$ of the cell surface, oriented 
towards the outside, will be equal to, respectively:
$$
\d\Pi_i=T_{ij}\d S_j\quad{\rm and}\quad M_i=\epsilon_{ijk}x_j\d\Pi_k.
\eqno({\rm A}2)
$$
From here, the total torque acting of the spherical surface in exam, can be expressed, 
after a few manipulations, in the form:
$$
{\bf M}=\mu\nu x\int\d\Omega_x\{[\x\times\nabla](\x\cdot\v)
+[\x\times(\x\cdot\nabla)\v]-[\x\times\v]\}.
\eqno({\rm A}3)
$$
The decomposition in vector spherical harmonics is used again:
$$
\v=v^{\rm s}\x Y+v^{\rm e}\nabla Y+v^{\rm m}[\x\times\nabla]Y,
\eqno({\rm A}4)
$$
where the notation: $v^{\rm sem}Y\equiv\sum_{lm}v^{\rm sem}_{lm}Y_{lm}$ is used. 

Substituting back into Eqn. (A3), one obtains the result that only the $v^{\rm m}$ 
components contribute to ${\bf M}$:
$$
{\bf M}=-\mu\nu x^5(v^{\rm m}/x)'\int\d\Omega_x\nabla Y
\eqno({\rm A}5)
$$
and, remembering the correspondence between spherical harmonics and irreducible tensors, it 
appears that only the $l=1$ components survive. The no torque condition takes then the form, 
from Eqn. (10):
$$
c_{l=1}=0.
\eqno({\rm A}6)
$$
Turning to the requirement of energy conservation and steady state, the equation for the
balance between dissipation and external work can be written in the form:
$$
W=\int\v_B\cdot\d({\bf\Pi}'-{\bf\Pi})=0,
\eqno({\rm A}7)
$$
where $W$ is the membrane dissipation, taken equal to zero, and $\d{\bf\Pi}'$ is the force 
with which the fluid 
inside the cell acts on the infinitesimal element $\d S$ of the membrane. If the vesicle
is perfectly spherical, it will rotate as a whole, so that $W=0$, $\d{\bf\Pi}'=0$ and 
Eqn. (A7) will coincide with the no torque condition ${\bf M}=0$. In fact, in this case
only one parameter, $v_B$, remains to be calculated.

For small non-zero $\epsilon$, $W$ is a linear functional of the external velocity field 
and can be decomposed into a contribution due to the strain $S$ and another to the vorticity 
$\omega$ of $\bar\v$:
$$
W=W^S(\barthe,\epsilon,S)+W^\omega(\barthe,\epsilon,\omega).
\eqno({\rm A}8)
$$
If $\omega=0$ the cell will tend to align with the strain axes of $\bar\v$, i.e.: $\barthe=\pm
{\pi\over 4}$; therefore:
$$
W^S(\pm{\pi\over 4},\epsilon,S)=W^S(\barthe,0,S)=0
\eqno({\rm A}9)
$$ 
and the first nonzero derivative of $W^S$ is: $\partial_\barthe \partial_\epsilon
W^S(\pm{\pi\over 4},0,S)$. Turning to the vortical component, writing:
$$
M=\int\d S A_M\quad{\rm and}\quad W^\omega=\int\d S A_W,
\eqno({\rm A}10)
$$ 
one observes that:
$$
A_W=(c+\O(\epsilon))A_M+\O(\epsilon^2).
\eqno({\rm A}11)
$$
Thus, if the no torque condition $M=0$ is satisfied, one has: $W^\omega(\barthe,\epsilon,\omega)
={\epsilon^2\over 2}\partial_\epsilon^2W^\omega(\barthe,0,\omega)+\O(\epsilon^3)$, and one finds
from Eqn. (A7):
$$
\barthe=\pm{\pi\over 4}-{\epsilon\over 2}{\partial_\epsilon^2W^\omega\over
\partial_\barthe \partial\epsilon W^S}+\O(\epsilon^2).
\eqno({\rm A}12)
$$ 
Hence, almost spherical vesicles tend to carry on a tank-treading motion, with the symmetry
axis alligned with one of the strain axes of the external flow.

\vskip 2cm

\end{document}